# Even-odd effect in multifragmentation products: the footprints of evaporation

*M.V. Ricciardi[1], K.-H. Schmidt[1], A. Kelić-Heil[1], Paolo Napolitani[1,2]*

1 GSI, 64291 Darmstadt, Germany
2 CNRS IN2P3, 91406 Orsay, France

**Abstract**
The analysis of experimental production cross-sections of intermediate-mass fragments (IMF) of several nuclear reactions at relativistic energy, measured at the FRS, GSI Darmstadt, revealed a very strong and complex even-odd staggering. The origin of this effect is related to the condensation process of hot nuclei while cooling down by evaporation. The characteristics of the staggering correlate strongly with the lowest particle separation energy of the final experimentally observed nuclei, but not with the binding energy. The study confirms the important role of the de-excitation process in multifragmentation reactions, and indicates that sequential decay strongly influences the yields of IMF, which are often used to extract information on the nature of nuclear reactions at high energies.

## 1 Introduction

In the last decades, a lot of effort has been invested to determine fundamental properties of hot nuclear matter. From the experimental side, successful studies gave indication of the liquid-gas phase transition in nuclear matter by the analysis of observables such as bimodality behaviours of the order parameters of the phase transition [1]. In parallel, the understanding of the thermal characteristics of a finite nucleus largely profited from the experimental determination of the caloric curve, which was achieved by the study of different properties of multifragmentation products formed in high-energy nuclear reactions. In particular, from the characteristics of the intermediate-mass fragments (IMF) produced in such reactions energy and temperature of the hot nucleus are deduced. While the evaluation of the excitation energy per nucleon ($E*/A$) is considered to be under control [2], larger problems are encountered in the measurement on the nuclear temperature $T$ [3]. Often, nuclear temperature was deduced from studying the thermal characteristics at freeze-out: such a situation corresponds to the instant of the collision process when the hot fragments which are formed do not exchange any more nuclear interactions. It is also assumed that the fragment configuration at freeze-out is the result of a chemical and thermal equilibration process which explored the coexistence of a dense (liquid) and diluted (gas) phase. From this time on, the evolution of the system is no more ruled by the transport dynamics but it is only governed by the Coulomb propagation and secondary decay. For this reason, in the following we may use the term "pre-fragment" to underline that the composite fragments formed in the collision process could be hot and therefore it could not coincide with the experimentally-observed final cold fragments, due to the loss of mass by evaporation during the de-excitation process.

Thermometers which extract $T$ from the measured yields of IMFs assume that the population at freeze-out follows a Boltzmann distribution. The probability of forming a pre-fragment $(A,Z)$ in a heat bath with temperature $T$ is proportional to [4]:

$$Y(A,Z) \propto \sum_i g_i e^{-E_i/T} \cdot e^{-\mu/T} \qquad (1)$$

where the sum extends over all possible energy states; $g_i$ is the degeneracy. The second factor comes from the condition of chemical equilibrium. The chemical potential $\mu$ depends on the binding energy $B(A,Z)$ of the pre-fragment: $\mu=\mu(B(A,Z))$. The Boltzmann factor $e^{-E_i/T}$ acts as a weighting factor, such that the lower is the energy $E_i$ of the state, the higher is the probability that the pre-fragment will be in that state. Since the excitation energies of very light pre-fragments are not significantly lower than $T$, it is assumed [4] that the population of light pre-fragments at freeze-out is well represented by the ground-state population. Under this assumption, the summation in Eq. (1) is limited to the ground state, whose energy $E_{gs}$ is zero; the first factor in Eq. (1) reduces to the ground-state spin $g_{gs}$, and the yield $Y(A,Z)$ is proportional to $e^{-\mu(B(A,Z))/T}$. If this were the case, the structural effects of the population at freeze-out should reflect those in the binding energies, among which those due to the pairing interaction. If we assume also that no further changes in the energy of the freeze-out population occur (for instance due to dynamical effects) we can expect that structural effects of the population at freeze-out reflect those of the measured final yields.

As a fundamentally different case, we could imagine the scenario where the pre-fragments are found in excited states with a high probability. If a dominant fraction is found with excitation energies above the critical pairing energy no structural effects related to pairing will be present in the formation cross sections of the pre-fragments.

In this work, we will test a very specific experimental signature, the even-odd structure in the yields of light nuclei, with the aim of discriminating between the two cases described above.

## 2    Experimental results

In 2004, the analysis of experimental production cross-sections of heavy IMFs ($Z \geq 7$) produced in the reaction $^{238}$U+Ti at 1 $A$ GeV revealed a very strong and complex even-odd staggering [5]. The experiment was performed using the high-resolution magnetic spectrometer FRagment Separator (FRS), at GSI. The produced nuclei were fully identified in mass and atomic number over an extended area of the chart of the nuclides, and kinematically separated to disentangle the different contributing reaction mechanisms (namely, fragmentation and fission). In the meantime, a large amount of new experiments devoted to measuring production cross-sections of fragments formed in nuclear reactions at high energy were performed at the FRS, GSI [6]. Similar experiments were performed at lower energies with the A1900 spectrometer at MSU, Michigan State, USA [7,8,9]. Recently, we analyzed the results from GSI and MSU experiments [10], and found that the cross sections of the produced nuclei also appear to be modulated by the same complex and very strong even-odd structure, previously observed in 2004. The new systematic analysis [10] of the even-odd effect over a large range of nuclear reactions (fission, spallation, fragmentation, multifragmentation) and nuclear systems (from iron till uranium) confirms the expected behaviors for the relatively heavy ($Z \geq 10$) residual nuclei produced in rather violent collisions: The structural effects cannot be attributed to the surviving of nuclear structure of the colliding nuclei, like it was interpreted in low-energy fission [11,12][1]; the structures appear as the result of the condensation process of hot nuclei while cooling down in the evaporation process [5].

---
[1] A more recent work concerning low-energy fission [12] shows that only in symmetric splits the even-odd effect in the yields can be fully attributed to the survival of pairing correlation, while in case of asymmetric splits, a much stronger contribution to the even-odd staggering comes from an *energy sorting* mechanism which *cools down* the light fragment.

Here, we want to focus on the reaction $^{56}$Fe+Ti at 1 $A$ GeV [13], where it was possible to extend the measurement of production cross sections down to lithium isotopes (Z=3). The experiment was performed at the FRS, at GSI, Darmstadt. As in the previous experiments, residual nuclei were fully identified in mass and atomic number and their production cross sections were measured. The longitudinal velocity of the fragments was measured with great precision exploiting the high resolution of the magnetic spectrometer. The combined information on yields and velocities of the produced fragments allowed disentangling multifragmentation yields from binary-decay yields [14,15]. In the following, we refer exclusively to multifragmentation products. More details on the experimental technique and on the data analysis can be found in ref. [13]. In Fig.1, the data are presented according to the neutron excess $N$-$Z$. The production cross sections of the observed fragments, grouped according to this filter, reveal a complex structure. All even-mass nuclei (left panel) present a visible even-odd effect, which is particularly strong for $N=Z$ nuclei. Odd-mass nuclei (right panel) show a reversed even-odd effect with enhanced production of odd-$Z$ nuclei. This enhancement is stronger for nuclei with larger values of $N$-$Z$. However, for nuclei with $N$-$Z$=1 the reversed even-odd effect vanishes at about $Z$=16, and an enhanced production of even-$Z$ nuclei can again be observed for $Z \geq 16$. Contrary to other neutron-rich chains, the neutron-deficient chain $N=Z$-1 shows a "typical" even-odd effect, i.e. an enhanced production of even-$Z$ nuclei.

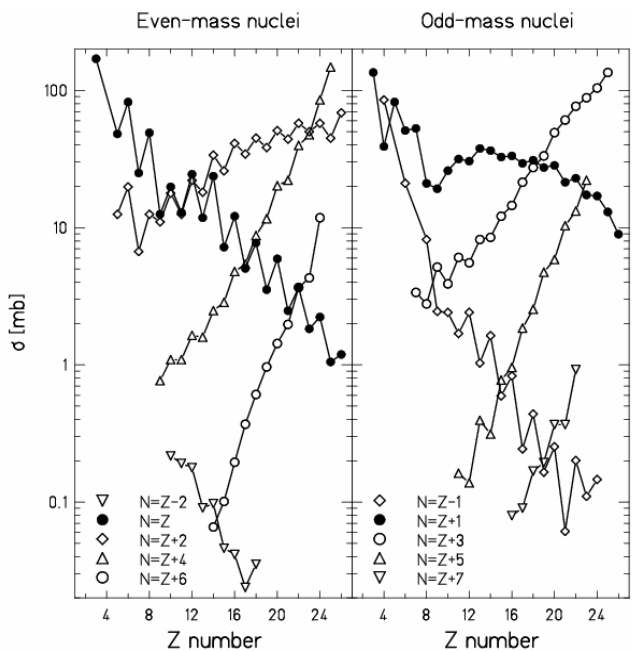

**Fig. 1:** Measured fragmentation cross sections of the residues from the reaction $^{56}$Fe + Ti, 1 $A$ GeV [13,16]. The particle-unstable nuclei $^{8}$Be and $^{9}$B were not measured. The data are grouped in sequences of nuclei with given value of $N$-$Z$.

## 3  Investigation of the population of fragments at freeze-out

By analysing the longitudinal velocity of the residual nuclei, P. Napolitani *et al.* [13] showed that the lightest fragments of Fig. 1 are produced in multifragmentation events. In this work, we assume that the freeze-out configuration in the multifragmentation process results from a situation of thermal and chemical equilibrium. This assumption is probably justified for nucleus-nucleus reaction at 1 GeV, since dynamical effects should not play a major role.

Hereafter, we want to show that the analysis of the even-odd staggering in the production cross-sections of these fragments can gives us very specific information on the nature of the population of fragments at freeze-out. We will consider two possible scenarios at freeze-out.

In the first scenario, the IMFs are essentially produced in their ground state; the yields are governed by Boltzmann statistics under certain conditions. In this case, it is not the phase space provided by the IMFs, which determines the yield, because each fragment offers only one state. It is the population probability of just the ground state of each IMF as part of all the possible states of the total system, which is decisive. In other words, the probability for the production of one or the other IMF is given by the phase space of the rest, which is left over when the IMF is produced. It is high if the energy left for the rest is high; it is low if the energy for the rest is low. This implies that the essential parameter is the binding energy of the respective IMF: $Y(A,Z) \propto e^{-B(A,Z)/T}$.

In the second scenario, the intermediate-mass pre-fragments have excitation energies above the critical pairing energy. In this case, the population of pre-fragments cannot show any even-odd staggering, because there is no pairing interaction above the critical energy. Since the pairing critical energy is also above the threshold for particle decay in the majority of cases, the pre-fragments will decay by evaporation. The population of pre-fragments will change during the de-excitation process, but it will remain always smooth as long as the excitation energy remains above the pairing critical energy. Only when the excitation energy drops eventually below the pairing critical energy, pairing correlations are established and even-odd structures may appear.

### 3.1 Even-odd staggering in the binding energy

The even-odd staggering in the binding energy was deeply investigated since several decades. We want just to recall here the main characteristics, as explained by Myers and Swiatecki in ref. [17]. Based on the idea that the interaction energy between two fermions will be greater when the two interacting densities have identical (congruent) nodal structures, compared to the case of two uncorrelated densities, Myers and Swiatecki calculated with simple combinatory the number of pairs with identical spatial wave-functions. As a result, they got that extra binding associated with the presence of congruent pairs is expressed by:

$$-\frac{3}{2}A + |N - Z| + \delta \qquad (2)$$

with $\delta = 0$ for even-even nuclei, $\delta = \frac{1}{2}$ for odd-even or even-odd nuclei, $\delta = 1$ for odd-odd nuclei, and $\delta = 2$ for $N=Z=$odd nuclei, a class of particularly unbound nuclei. The interaction energy between a pair of nucleons interacting by short-range forces is inversely proportional to the volume – itself approximately proportional to $A$ – in which the nucleons are confined. It follows that the extra binding associated with the presence of congruent pairs is:

$$-\frac{3}{2} + |I| + \frac{\delta}{A} \quad \text{with} \quad I = \frac{N-Z}{A} \qquad (3)$$

The quantity $C = -3/2 + |I|$ is the "congruence energy" and $\delta/A$ is the pairing correction. The congruence energy is often parameterized as $C = -a + b \cdot |I|$ ($a \sim 7$ MeV and $b \sim 42$ MeV), where $b \cdot |I|$ is the Wigner term. Thus, the staggering in binding energy can come only from the staggering of $\delta$. In Fig. 2, the experimental staggering in the binding energy is presented for four sequences of nuclei with constant neutron-excess: $N=Z-1$, $N=Z$, $N=Z+1$, $N=Z+2$. The binding energy staggering is given by the difference between the tabulated experimental binding energies from Audi and Wapstra [18] and the liquid drop binding energies, calculated according to the model of ref. [19]. In the calculated liquid drop binding energies nor pairing nor shell contributions were included. Please note that the long-range fluctuation is related to shell effects (not considered in the liquid-drop formula).

Although the formulation presented in ref. [14] is not precise on a quantitative level (for example, the magnitude of the staggering for N=Z odd nuclei is too strong), the staggering in the experimental binding energy is qualitatively consistent with the prescription of ref. [17]. The $N=Z$ and $N=Z+2$ chains are made of even-even and odd-odd nuclei alternating; the extra binding energy associated with pairing alternates between $\delta=0$ and $\delta=2$ (for the $N=Z$ chain) and between $\delta=0$ and $\delta=1$ (for the $N=Z+2$ chain), producing a strong even-odd staggering. The $N=Z-1$ and $N=Z+1$ chains are made of even-odd and odd-even nuclei alternating; the extra binding energy associated with pairing is given by $\delta=½$ for both type of nuclei, producing no even-odd staggering.

Thus, we can conclude that the staggering in the binding energies cannot be responsible for the observed even-odd staggering in the final yields. Consequently, the first scenario described above seems not to be realistic.

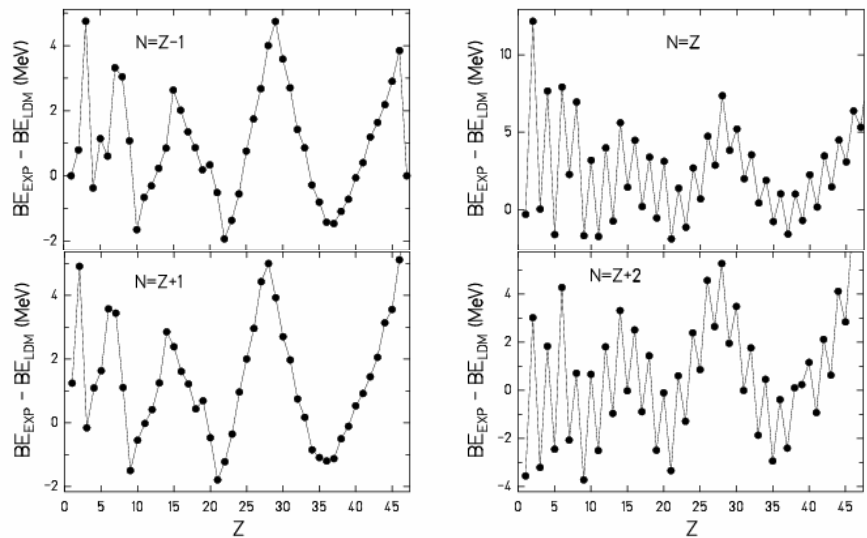

**Fig. 2:** Binding energy staggering given by the difference between the tabulated experimental binding energies from [18] and the liquid drop binding energies [19].

### 3.2 Even-odd effect in the remnants of an evaporation chain

We want to discuss now which kind of staggering we have to expect in the cold final products due to the influence of an evaporation cascade starting from hot pre-fragments. In this case, after their formation, the initial pre-fragments are definitely free of any even-odd structure, because there is no pairing above the critical energy. There are a large number of states available for the direct production. The number of available states from certain total excitation energy of the decaying system is essentially determined by a mostly structureless level density above a fictive liquid-drop ground state. Some influence of shell effects is still possible. This produces an essentially smooth 2-dimensional distribution in $N$ and $Z$ on the chart of the nuclides with an excitation-energy distribution, which varies smoothly as a function of $N$ and $Z$. This structureless cloud rains down in excitation energy and in nucleon number due to evaporation. Since in most cases one particle dominates and the variation of the kinetic energy of the emitted particle is small compared to the binding energy, this process is *almost* deterministic. In this way, a well-defined 3-dimensional subspace of the $N,Z,E^*$ initial population ends up in a certain fragment. As a consequence, the final yields will be modulated by the range of excitation energy below the lowest particle threshold: There is a fine structure, proportional to the fluctuations of the lowest particle threshold. This is an old idea of J. Hüfner, C. Sander and G. Wolschin presented in ref. [20] (see in particular Fig. 1 from ref. [20]), where they wrote: "... the decay is strongly determined by threshold effects. The production cross sections are determined by how much excitation cross-section there is in a given interval between two thresholds, *not* by how the energy is precisely distributed over the available degrees of freedom". This idea is at the base of the de-excitation model of X. Campi and J. Hüfner [21], later modified and improved by J. J. Gaimard and K. H. Schmidt [22], where the evaporation stage is treated in a macroscopic way on the basis of a master equation which leads to a diffusion equation. There, it is written: "The mass yield curve $\sigma(A)$ is directly related to the pre-fragment distribution because of conservation of probability and energy, essentially".

We can clarify the above concept, also observing the individual stories of pre-fragments. During the evaporation process the population of pre-fragments is smooth as long as the excitation is high, because there cannot be pairing interactions. When the excitation energy of a given pre-fragment falls below the critical energy for pairing (about 10 MeV) structural effects due to pairing can be restored. Normally, with 10 MeV of excitation energy, only one evaporation step is possible. Hereafter we assume that the even-odd effect is decided *in the last evaporation step*. We assume further that in the last evaporation step *only the emission of a proton or of a neutron occurs*. The evaporation of heavier particles is a more infrequent decay process, which is energetically more expensive, and normally takes place at the beginning of the evaporation cascade where higher excitation energy is available. In Fig. 3, as an example, we depicted a possible last-evaporation-step. The final fragment ($N,Z$) can be produced either by the decay of a proton from the mother nucleus ($N, Z+1$) or by the decay of a neutron from the mother nucleus ($N+1, Z$). In the example, the proton separation energy $S_p$ of the final fragment ($N,Z$) is lower than its neutron separation energy $S_n$. Four cases are possible:

1) The excitation energy of the mother nucleus is represented by the green dashed-dotted line. Both possible mother nuclei can decay into ($N,Z$). The decay is energetically possible, and could happen; however the daughter nucleus ($N,Z$) would have excitation energy above its particle thresholds and would decay further by neutron or proton emission. So, it would not be the last evaporation step and ($N,Z$) would not be our observed final fragment.

2) The excitation energy of the mother nucleus is represented by the red dotted line. Both possible mother nuclei can decay into ($N,Z$). The decay is energetically possible, and could happen; however the daughter nucleus ($N,Z$) would have excitation energy above its proton separation energy threshold and would decay further by proton emission. So, it would not be the last evaporation step and ($N,Z$) would not be our observed final fragment.

3) The excitation energy of the mother nucleus is represented by the blue dashed line. Both possible mother nuclei can decay into ($N,Z$). The decay is energetically possible, and could happen. The daughter nucleus ($N,Z$) would have excitation energy above below its lowest particle threshold and would decay further only by gamma emission. So, ($N,Z$) would be our observed final fragment.

4) If the excitation energy of the twp possible mother nuclei is below their proton or neutron separation energy, they could not decay into ($N,Z$), which would not be our observed final fragment.

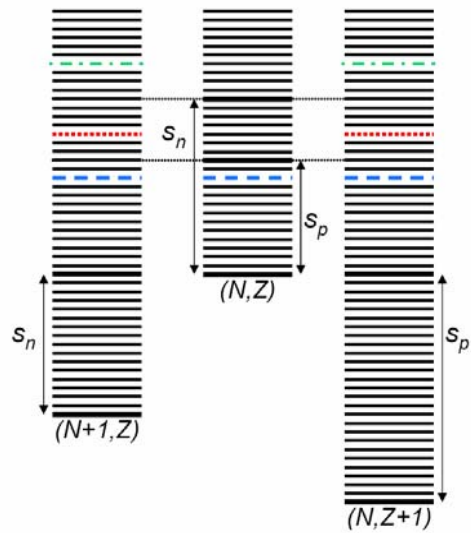

**Fig. 3:** Schematic view of the last evaporation step for the final fragment (*N,Z*) originated from the possible decay of two mother nuclei: (*N+1,Z*) and (*N,Z+1*).

We can consider now, in Fig. 4, the last evaporation step for the 4 possible types of final fragments: odd-odd, even-odd, odd-even, even-even. In the figure, the blue dashed line represents the lowest particle separation energy of the final fragment. The reasoning done for the example presented in Fig. 3 can be applied also in all other cases: in the last evaporation step, the mother nucleus must have excitation energy $E^*$ between the lower dotted red line and the upper dashed blue line. In reality, the mother nucleus can have $E^*$ a little bit above the upper dashed blue line and still decay in the daughter nucleus, because some energy would go in the kinetic energy of the emitted particle, but this quantity is small and in first approximation we can neglect it. We can conclude that the two lines define the "energy range" occupied by the mother nucleus: if the energy of the mother nucleus is between the ground state and particle separation energy of the daughter nucleus, then it will decay into the daughter nucleus. The same reasoning is valid also for the case of an even-even final fragment, whose first excited state is well separate from the ground state. If the energy of the mother nucleus is above its particle separation energy (lower red dotted line) and below the dashed-dotted line, it will decay into the ground state of the daughter nucleus; this does not make any difference in terms of final yield. Here we must point out that things go differently for heavy nuclei, where gamma emission becomes a competitive channel, as discussed in ref. [5].

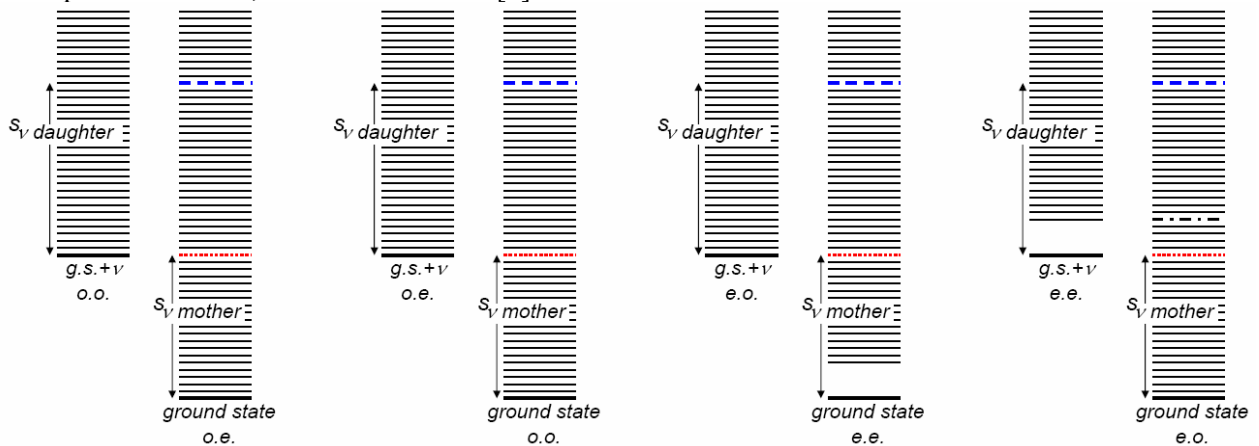

**Fig. 4:** Schematic view of the last evaporation step. Four possible cases are shown, which represent four possible transition from a compound nucleus (the mother nucleus, always to the right) to a final nucleus (the daughter nucleus, always to the left: from odd-even to odd-odd, from odd-odd to odd-even or even-odd, from even-even to odd-even or even-odd, and from even-odd to even-even. The label ν stands for either n (neutron) or p (proton). The y scales represent the absolute energy relative to the ground states of the mother nuclei in arbitrary units and not to scale.

The important conclusion here is that the phase-space, which is related to the level density, is not anymore the relevant quantity in the last evaporation step, which can be considered quasi deterministic. It is the separation energy, which gives the range of excitation energy, which "catches" the evaporation flux in particle-stable states, in the sense of Campi and Hüfner. The relevant quantity for the staggering in the final yields is the "particle threshold", which we called "energy range", represented by the lowest value of the particle separation energy of the daughter nucleus.

In Fig. 5, we present on the chart of the nuclides the lowest particle separation energy for light nuclides, calculated as the minimum value between $S_n$, neutron separation energy, and $S_p$, proton separation energy, for each nuclide. The values of proton and neutron separation energies, $S_p$ and $S_n$, were taken from the compilation of Audi and Wapstra [18]. As stated before, it is the lowest particle separation energy which determines the final flux of phase space, and therefore, the staggering in the final yields. The interplay of the neutron and proton separation energies forms a kind of "fishbone pattern". Apart from the fine structure, the global variation of the threshold energy as a function of N-Z creates also an additional long-range modulation of the cross sections, which adds up to the influence of the N/Z of the initial pre-fragments and to the "direction" of the evaporation process on the chart of the nuclides. In other words, the decrease of the separation energy towards the driplines, implies a decrease of the production yields, which is observed indeed experimentally in the production cross sections.

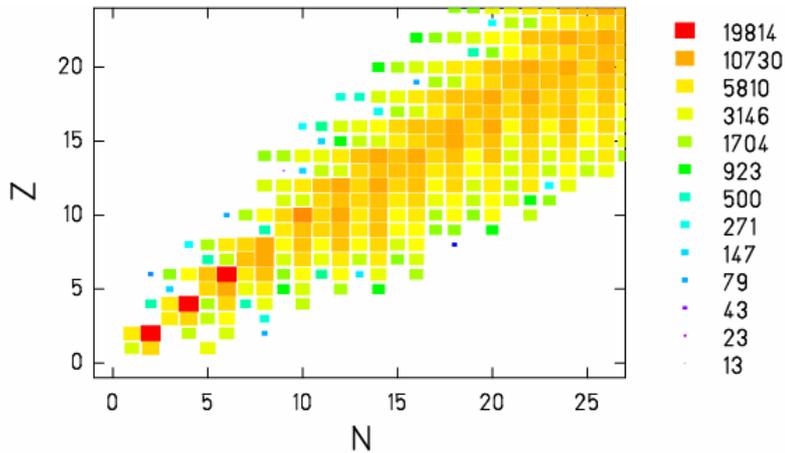

**Fig.5:** Lowest particle separation energy for each nuclide (keV). Values are taken from Ref. [18].

In reality, the proton separation energy should not be compared directly to the neutron separation energy because of the Coulomb barrier. $S_n$ and $S_p$ are approximately equal along the stability line. However, when we consider an evaporation process, the emitted particle has to overcome the Coulomb barrier. As a consequence, an excited nucleus close to the stability line will continue to evaporate neutrons till it reaches a certain position on the neutron-deficient side of the chart of the nuclides. This is why the "evaporation corridor" [23] is defined by the "attractor line" [24] and not by the stability line. A more precise evaluation of the "energy range" should take into account the Coulomb barrier. However, to calculate the increase of $S_p$ to get the "effective $S_p$" is not trivial because for proton emission tunneling through the barrier is not at all negligible. On the other hand, $S_p$ is given by the mass difference of the two close nuclei, so at infinite distance, *not* above the Coulomb barrier. So, to increase $S_p$ by an amount equal to a coulomb barrier would also be wrong. Anyhow, Coulomb effects are small for the light nuclei considered here, and might be neglected in the first step. In addition, the effects of the Coulomb barrier are smooth, and do not introduce any additional staggering, so they do not affect our qualitative description. We must also point out that for a few light nuclei $S_\alpha$, the separation energy for the emission of alpha particle ($S_\alpha = -Q(\alpha)$, where $Q(\alpha) = M(A,Z) - M(A-4,Z-2) - {}^4He$), is lower than $S_n$ and $S_p$, and therefore should be taken as representative of the particle threshold. However, as for the proton emission, one should take into account the effects of the Coulomb barrier. We decided to neglect these few cases. In conclusion, we think that the lowest neutron and proton particle separation energy is a good representative of the particle threshold and therefore of the "energy range", at least to understand qualitatively the staggering phenomenon.

A remark concerning the importance of the level density is due. We pointed out that in the last evaporation step the lowest neutron and proton particle separation energy well represents the particle threshold. In turn, it is the particle threshold which determines the final flux of phase space, and therefore, the staggering in the final yields. In this sense, a detailed knowledge on level density is not needed to understand qualitatively the even-odd staggering in the yield. However, for a precise quantitative description of the final yields, one has to consider that in light nuclei the emission of neutrons and protons competes with other decay channels in specific cases. Therefore, a statistical treatment also of the last evaporation step is more appropriate for a quantitative description.

## 4    Origin of the even-odd effect in light fragments

We have now the information to understand the origin of the even-odd effect in the yields of the reaction ${}^{56}$Fe+Ti at 1 $A$ GeV, presented in Fig.1. The lowest particle threshold presented in Fig. 5 builds up a "fishbone" pattern, which emphasizes a strong and complex even-odd staggering. Fig. 5 shows that along the chain $N=Z$, even-even nuclei show the highest energy range. Along the chain $N=Z+1$, it is the energy range of $Z=$odd nuclei which is enhanced. On the contrary, along the chain $N=Z-1$, it is the energy range of $Z=$even nuclei which is enhanced. These features are fully consistent with the staggering in the experimental yields. On the other hand, the results presented in Fig. 2 indicate that the binding energy of nuclei with N=Z do not show any staggering, contrary to what observed in the experimental yields of Fig. 1.

In Fig. 6, we focus on two sequences: the $N=Z$ chain and the $N=Z+1$ chain. These are the only two chains of fragments that extend down to lithium isotopes. The experimental cross sections (expressed in mb) are compared with the biding-energy staggering (expressed in MeV) and with the lowest particle separation energies (expressed in MeV).

We want to make a short comment regarding the possibility that the degeneracy factor introduces an additional staggering. Under the assumption that only the ground state is populated at freeze-out, we have to consider only the ground-state spin. For the chain N=Z+1 the ground-state spin from ${}^7$Li to ${}^{17}$O is 3/2, 3/2, 3/2, 1/2, 1/2, 5/2 respectively; thus it does not stagger. For the chain N=Z the ground-state spin from ${}^6$Li to ${}^{16}$O is 1, 0, 3, 0, 1, 0 respectively. Here, the staggering of the ground-state spin would enhance the production of even-$Z$ nuclei, a fact which is not observed in the experimental data, indicating that evaporation process is dominant and hides any possible indication of the original staggering after break-up.

The results indicate clearly that even the lightest IMFs are predominantly the end-products of an evaporation process. At freeze-out, the population of IMFs in excited states is large enough to originate a sequential evaporation cascade. The end products of this evaporation cascade sum up to the population of nuclei produced in their ground state (a phenomenon called "side feeding"). If the contribution from side feeding would be negligible, the even-odd staggering would still keep

(although smoothed) the characteristic of the ground-state population. This fact is not observed. We conclude that the first scenario proposed in section 3 is not realistic, while the data seem to hint for the validity of the second scenario.

Is difficult to establish, even for very light IMFs such as lithium isotopes ($Z=3$), how much their final measured yields reflect the original ground-state population at freeze-out. The data indicate that side-feeding from heavier IMFs is dominant; in addition we cannot exclude that the population of lithium isotopes at freeze-out is itself partly in higher energy states which further decay by particle or gamma emission.

It was pointed out that the fluctuations of the temperature measured with several isotope thermometers appear to originate from structural effects in the secondary decay process [25]. To express it more properly, the structural effects modulate the thresholds inside which the evaporation flux is "caught" in particle-stable states, in the sense of Campi and Hüfner [20, 21]; it is the evaporation process that introduces differences between the measurements of temperature with several isotope thermometers.

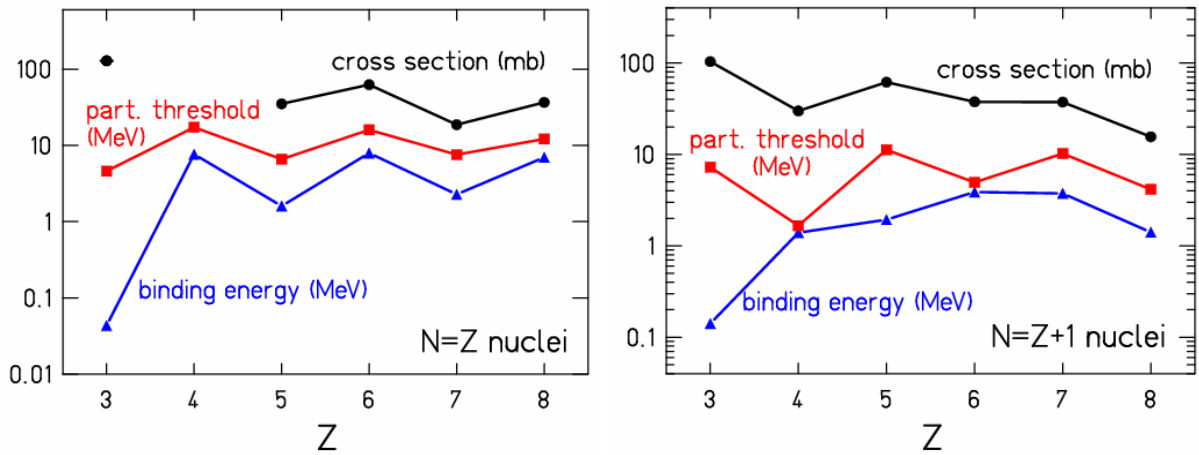

**Fig.6:** Results for the sequences of nuclei with $N=Z$ (left) and $N-Z=1$ (right). Dots: Measured fragmentation cross sections of the residues from the reaction $^{56}$Fe + Ti, 1 $A$ GeV [13] ($^{8}$Be is not bound, thus it cannot be measured). Squares: Lowest particle separation energy (particle threshold). Triangles: Extra binding energy due to pairing.

## 5 Conclusions

We have studied the origin of the even-odd staggering observed in the production cross-section from high-energy reactions. We concluded that the staggering in the final yields is determined to great extent in the last evaporation step. The last evaporation step can be considered quasi deterministic. The phase-space, which is related to the level density, is not the relevant quantity anymore in the last evaporation step: It is the separation energy, which gives the range of excitation energy, which "catches" the evaporation flux in particle-stable states, in the sense of Campi and Hüfner [20, 21]. Therefore, the relevant quantity for the staggering in the final yields is the lowest particle threshold. An excellent description of the complex even-odd staggering observed experimentally was obtained considering as lowest particle threshold the minimum value between proton and neutron separation energies. The observed fine structure in the yields of the fragments correlates with the proton and neutron separation energies. The structure is not consistent with the fluctuations of the binding energies.

This finding proves that the primary fragments produced in multifragmentation before secondary decay are mostly produced in excited states. The direct production in their ground state seems to be weak. We conclude that it is dangerous to extract fundamental properties of hot nuclear matter assuming that the final yield is represented directly by the Boltzmann factor. It has already been recognized previously that the determination of the freeze-out temperature with the isotopic yield thermometer might be disturbed by the evaporation process or sequential decay. Several authors investigated this problem and proposed suitable corrections to this effect [25,26]. Our analysis opens a new view on this problem by showing –in a totally model-independent way– that the fine structure observed in the final yields can fully be explained by the evaporation process. That means that this specific feature does not show any noticeable influence of the Boltzmann factor, which is behind the isotopic thermometer method. Quantitative conclusions on the validity of the isotopic thermometer method are beyond the scope of this work.

## Acknowledgments

We acknowledge the financial support of the European Community under the EUROTRANS Integrated Project FI6W-CT-2004-516520. Fruitful discussions with Francesca Gulminelli, Mauro Bruno, and Plamen Boutachkov are gratefully acknowledged.